\documentclass[article,12pt,numbers]{elsarticle}

\usepackage{amssymb}
\usepackage{amsmath}
\usepackage{graphicx}
\usepackage{anyfontsize}
\usepackage{xcolor}
\usepackage{steinmetz}

\graphicspath{{./images/}}

\journal{Science Bulletin}

\makeatletter
\let\@afterindenttrue\@afterindentfalse
\makeatother

\begin{document}

\begin{frontmatter}

\title{Topological surface phonons modulate thermal transport in semiconductor thin films}



\author[1,2]{Zhe Su\fnref{eq1}}
\author[1,2]{Shuoran Song\fnref{eq1}}

\author[2,3]{Qi Wang\corref{cor1}}

\author[1,2,3]{Jian-Hua Jiang\corref{cor1}}

\affiliation[1]{organization={School of Physical Sciences, University of Science and Technology of China},%
            city={Anhui},
            country={China}}

\affiliation[2]{organization={Suzhou Institute for Advanced Research, University of Science and Technology of China},%
            city={Jiangsu},
            country={China}}

\affiliation[3]{organization={School of Biomedical Engineering, Division of Life Sciences and Medicine, University of Science and Technology of China},%
            city={Anhui},
            country={China}}

\cortext[cor1]{Corresponding authors.}

\fntext[eq1]{Zhe Su and Shuoran Song contributed equally to this study.}

\begin{abstract}
While phonon topology in crystalline solids has been extensively studied, its influence on thermal transport-especially in nanostructures-remains elusive. Here, by combining first-principles-based machine learning potentials with the phonon Boltzmann transport equation and molecular dynamics simulations, we systematically investigate the role of topological surface phonons in the in-plane thermal transport of semiconductor thin films (Si, 4\textit{H}-SiC, and \textit{c}-BN). These topological surface phonons, originating from nontrivial acoustic phonon nodal lines, not only serve as key scattering channels for dominant acoustic phonons but also contribute substantially to the overall thermal conductivity. Remarkably, for these thin semiconductor films below 10 nm this contribution can be as large as over 30$\%$ of the in-plane thermal conductivity at 300 K, and the largest absolute contribution can reach 82 W/m-K, highlighting their significant role in nanoscale thermal transport in semiconductors. Furthermore, we demonstrate that both temperature and biaxial strain provide effective means to modulate this contribution. Our work establishes a direct link between topological surface phonons and nanoscale thermal transport, offering the first quantitative assessment of their role and paving the way for topology‑enabled thermal management in semiconductors.
\end{abstract}


\end{frontmatter}


\section{Introduction}\label{sec1}

Over the past decades, the notion of topological phases and their unprecedented properties has been studied for electronic and other systems, flourishing our understanding of materials~\cite{KMModelTopo, QuantumTopo_2007, SymIndicator,EleTopoCatalogue, EleTopoCatalogue_2, EleTopoCatalogue_3, TopoQuanChem_2017, CatalogPhonon, PhononBulkExp_2018, GrapheneTopoExp_2023, PhononTopoSearc}. A variety of three-dimensional topological insulators, topological Dirac points, Weyl points, nodal lines, and other novel topological states of electronic and phonon systems in solids have been investigated in both theory and experiments~\cite{wieder_topological_2021,zhang_new_2025}. The intriguing transport of topological surface electronic states has also been thoroughly investigated~\cite{surface_conduction_2012, tuning_surface_conduction_2015, nanoscale_ele_transport_2016, suface_diode_effect_2025}, but rarely in phonon systems~\cite{topo_phonon_thermoelectric_2024, topo_phonon_transport_bulk}. Phonons are important carriers of energy in thermal transport. The influence of phonon band topology and topological surface phonons on thermal transport in solid state materials remains rarely studied~\cite{topo_phonon_thermoelectric_2024,topo_phonon_transport_bulk}.

Unlike electronic systems which often have quantized topological transport properties (such as in the quantum Hall effects)~\cite{topo_ele_transport_surface, nanoscale_ele_transport_2016}, phonons do not have such transport properties as they do not have a Fermi level. All phonons, including those of both the low- and high- frequency as well as those in the bulk or at surfaces, participate in thermal transport through multiple-phonon scatterings, which brings difficulty in isolating and quantifying the contribution to thermal transport of topological surface phonons. Also, realistic surface conditions can be complicated due to, e.g., surface reconstructions~\cite{surface_reconst_Si, surface_reconst_GaN, dftb_Si_thin_films, Si_thin_film_exp, review_surface_cal_2018}, which pose computational challenges for theoretical calculations~\cite{dftb_Si_thin_films, Si_thin_film_exp, review_surface_cal_2018}. In addition, phonon anharmonicity is hard to incorporate in previous studies, which employed the S-matrix method to compute the phonon transmission spectrum of topological edge and corner states in a few perfect two-dimensional nanostructures~\cite{S_matrix_graphene,s_matrix_corner_state}. These facts make the study of topological phonons in thermal transport rather challenging. However, as the sizes of semiconductors in electronic devices shrink to nanoscales~\cite{nanoscale_thermal}, and taking into account that many solids (such as silicon~\cite{ubiquitous_silicon}) host topological phonon states, the investigation of topological surface phonons in thermal transport is highly anticipated, yet unfulfilled.

\begin{figure*}
    \centering
    \includegraphics[width=0.8\textwidth]{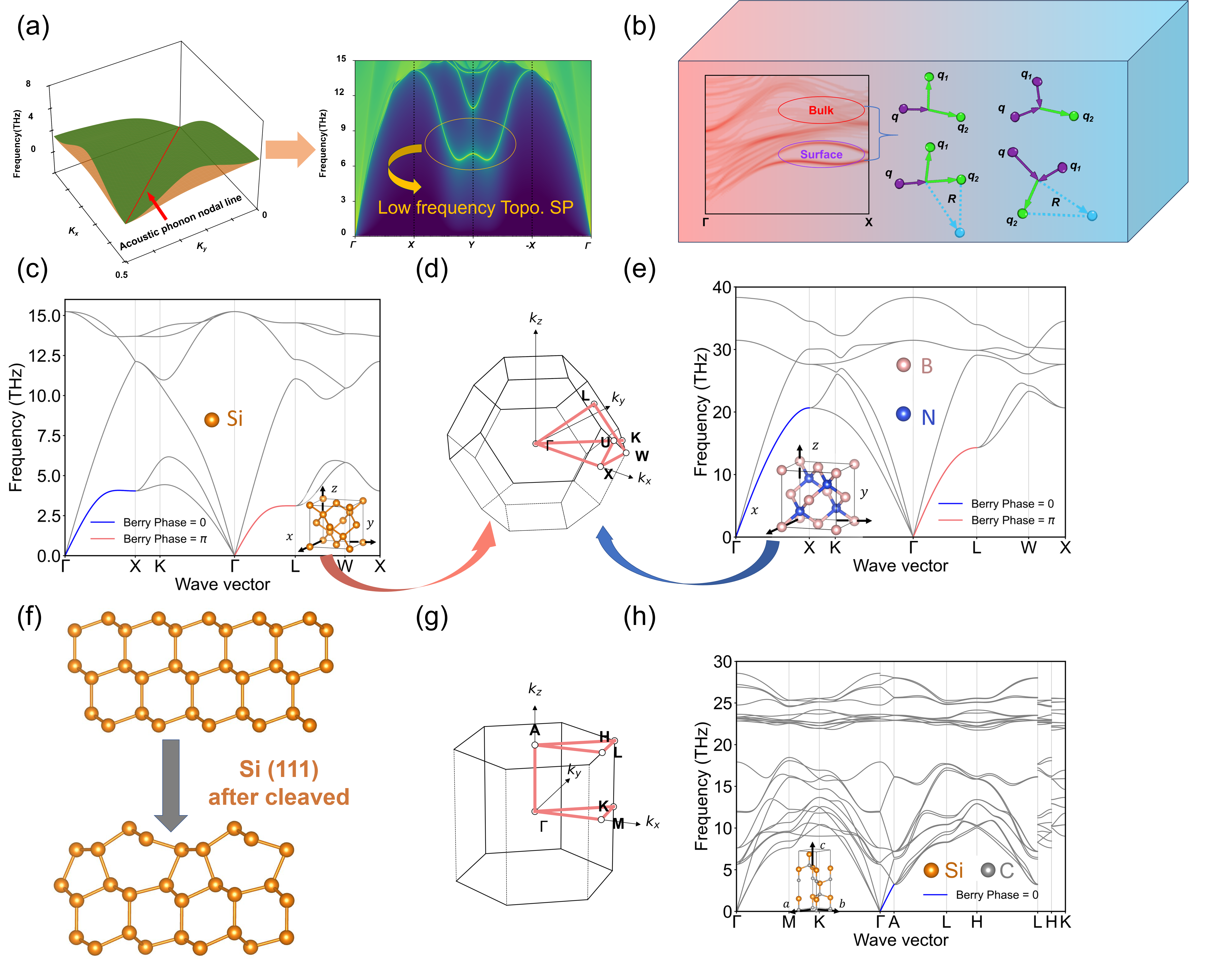}
    \caption{(a) Schematic of the acoustic phonon nodal line (left) and the corresponding low-frequency topological surface states (right); (b) Sketch of three-phonon scattering processes in a thin film, and both bulk and topological surface phonons can participate in phonon scattering. Phonon dispersion relations of bulk (c) Si, (e) 4\textit{c}-BN, and (h) \textit{4H}-SiC from first-principles calculations with their crystal structures. The Brillouin zones of bulk Si and \textit{c}-BN, and 4\textit{H}-SiC are shown in (d) and (g), respectively.}
    \label{fig-1}
\end{figure*}

\begin{figure*}
    \centering
    \includegraphics[width=0.65\textwidth]{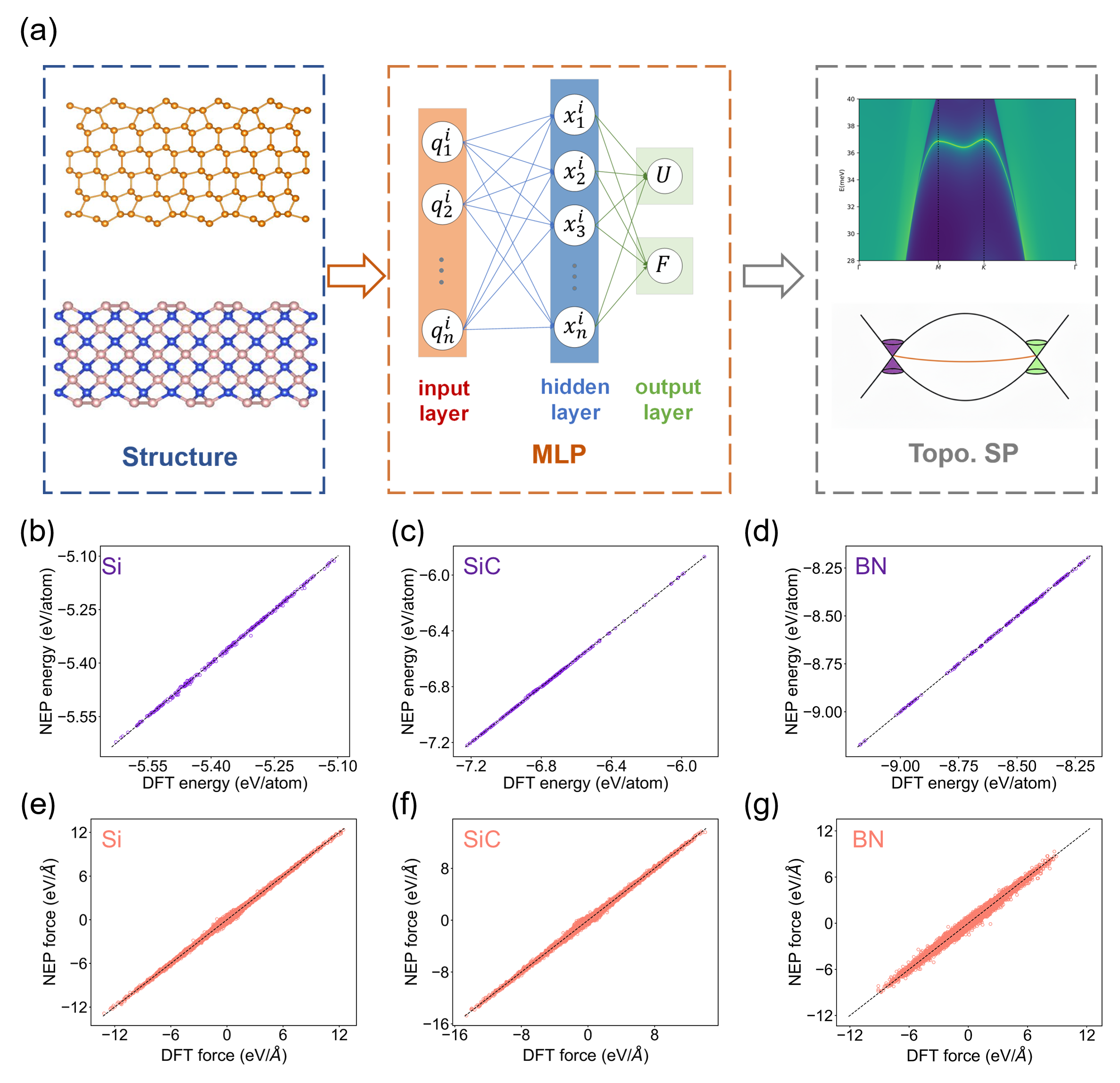}
    \caption{(a)Schematic of the process using MLPs to predict topological surface phonons (Topo. SPs). Comparison of energies and atomic forces of (b, e) Si, (c, f) 4\textit{H}-SiC, and (d, g) \textit{c}-BN calculated from the DFT and NEPs for the training sets.}
    \label{fig-2}
\end{figure*}

In this work, we fill this gap by combining first-principles-based machine learning potentials with the phonon Boltzmann transport equation (BTE) and molecular dynamics simulations to investigate the influence of topological surface phonons on thermal transport in thin films of three representative semiconductors: Si, 4\textit{H}-SiC, and \textit{c}-BN. Our study focuses on the low-frequency topological surface states associated with nontrivial acoustic phonon nodal lines and utilizes machine learning potentials to quantify their contributions to the overall thermal conductivity with first-principles accuracy. We demonstrate that low-frequency topological surface phonon states exist on various surfaces of the semiconductor thin films. For thin films with a thickness below 6 nm, the BTE results reveal that low-frequency phonons dominated by topological surface states provide additional phonon scattering channels for long-wavelength phonons. These low-frequency topological surface phonons also have relatively large phonon group velocity and long lifetimes, leading to a substantial contribution to in-plane thermal conductivity. As the film thickness decreases, the role of topological surface phonons in thermal transport becomes increasingly vital, accounting for up to 30$\%$ of the total in-plane thermal conductivity of certain thin films. Furthermore, We demonstrate that both temperature and strain can be effective approaches to tuning the contributions of topological surface phonons to the in-plane thermal conductivity of semiconductor films.

\section{Computational Methods}\label{sec2}

Phonon dispersion relations of bulk Si, 4\textit{H}-SiC, and \textit{c}-BN are calculated using the finite-displacement method and the harmonic interatomic force constants (IFCs) from the density functional theory (DFT) calculations via the phonopy package~\cite{phonopy}. The PBEsol exchange-correlation functional~\cite{PBEsol} for the projector-augmented-wave (PAW) method~\cite{PAW_dft} implemented in the Vienna ab initio simulation package (VASP)~\cite{vasp, kseq, plane_wave} is used for our DFT calculations for Si and \textit{c}-BN, while the revised PBE (revPBE) functional proposed by Zhang and Yang~\cite{revPBE} is used for 4\textit{H}-SiC. The harmonic IFCs and Green's function method with open-boundary conditions~\cite{surface_green_function} are adopted to obtain the various surface local density of states (LDOS) of Si, 4\textit{H}-SiC, and \textit{c}-BN following the method in Ref. \cite{ubiquitous_silicon}. Determination of other calculation parameters can be found in the Supporting Information.

The active learning strategy~\cite{nepactive} is used to select the training sets of different thin films of Si, 4\textit{H}-SiC, and \textit{c}-BN. Spin-orbit coupling (SOC) is considered only in the calculations of static energies and atomic forces for the 4\textit{H}-SiC configurations in the training set, as SOC can help maintain the stability of the (0001) surfaces~\cite{SiC_SOC}. 

The phonon BTE method as implemented in ShengBTE~\cite{shengbte} is used in computing the lattice thermal conductivity of Si, 4\textit{H}-SiC, and \textit{c}-BN thin films with the harmonic and third-order IFCs calculated by the neuroevolution potentials (NEPs). The mode-wise phonon group velocity, lifetime, and weighted phase space (WPS) can also be determined accordingly. For all semiconductor thin films, the homogeneous nonequilibrium molecular dynamics (HNEMD) method~\cite{hnemd} in the GPUMD software~\cite{gpumd_4.0} is also used to calculate the lattice thermal conductivity of thin films.

 \section{Results and Discussion}\label{sec3}
 \subsection{Phonon nodal lines and Zak's phase}

Low-frequency phonons are usually recognized as the dominant heat carriers in insulators and semiconductors~\cite{Si_dftb, GaN_dft, SiC_dft}. For instance, acoustic phonons contribute more than 90$\%$ to the total thermal conductivity of the bulk semiconductor silicon (Si) at room temperature, as revealed by first-principles calculations~\cite{Si_exp_dft}. Si, with a bulk thermal conductivity of 150 W/m-K~\cite{Si_kappa}, serves as a core semiconductor in modern electronics. 4\textit{H}-SiC, widely used in power electronics, exhibits a thermal conductivity of 490 W/m-K~\cite{SiC_kappa}, while \textit{c}-BN is another important semiconductor with an even higher thermal conductivity up to 1200 W/m-K~\cite{BN_kappa}. These materials belong to symmorphic and nonsymmorphic crystals with space groups $F\overline{d}3m$,  $P6_3mc$, and $F43m$ , respectively. 

We first identify whether the acoustic straight nodal lines of Si, 4\textit{H}-SiC, and \textit{c}-BN, are nontrivial. The lattice structures, phonon dispersion relations, and  Brillouin zones (BZs) of Si, 4\textit{H}-SiC, and \textit{c}-BN are displayed in Fig.~\ref{fig-1}. Nodal lines, serving as the one-dimensional form of band crossings in topological semimetals, can hold multiple promising physical properties. Many topological phonon nodal-line semimetals have been theoretically predicted and experimentally verified~\cite{ubiquitous_silicon,phonon_NL_3d,phonon_NL_MoB2,data_search_phonon_topo,phonon_NL_PRB,phonon_NL_review,phonon_NL_NS}. Owing to the bulk-boundary correspondence, nontrivial phonon nodal lines can give rise to unique topological surface states on specific surfaces, such as drumhead-like surface states~\cite{phonon_NL_review, ubiquitous_silicon}. Especially low-frequency topological surface phonons, which can exhibit large group velocities and long lifetimes, are potentially significant heat carriers. To date, however, the influence of these low-frequency topological surface phonons associated with acoustic nodal lines on phonon transport remains largely unexplored.

A two-band effective Hamiltonian model of  \textbf{\textit{k}}·\textbf{\textit{p}} theory can be generalized to describe the band crossing of two phonon branches~\cite{phonon_NL_PRB} in silicon,
\begin{equation}
H_e(\mathbf{q})=u(\mathbf{q})\sigma_++u^*(\mathbf{q})\sigma_-+g(\mathbf{q})\sigma_z . 
\end{equation}
Here, $H_e(\mathbf{q})$ is defined with the frequency reference set on a point of the nodal line, $u(\mathbf{q})$ and $g(\mathbf{q})$ are a complex and a real function, respectively, where the \textbf{\textit{q}} wave vector is in a plane perpendicular to the nodal line. $\sigma_{ \pm}(=\sigma_{x} \pm i \sigma_{y}$) and $\sigma_{0,x,y,z}$ are the Pauli matrices. In Si, the double degenerate phonon band along the $\Gamma$X path is protected by the little group $C_{4v}$, which contains two generators $C_{4z}$ and $M_x$. With the basis functions as {$F_x$, $F_y$}, the lowest order approximation of $H(\mathbf{q})$ is expressed as,
\begin{equation}
    H(\boldsymbol{q})=[c_1(q_+^2+q_-^2)+c_2q_+q_-]\sigma_z+ic_3(q_+^2-q_-^2)\sigma_+ +H.c.,
\end{equation}
where $c_i$ (\textit{i}=1, 2, 3) are real numbers and $q_\pm=q_x\pm iq_y$. Consequently, the leading-order term of the effective Hamiltonian for the acoustic straight nodal lines along the  path  $\Gamma$X is quadratic, yielding a Berry phase of $2\pi$, which is also consistent with our numerical calculations. In phonon systems, the Berry phase of a phonon branch \textit{n }can be defined as 
\begin{equation}
    \theta_{n}=\oint_{C}\!\mathbf{A}_{n\textrm{k}}\textrm{d}\mathbf{k}=\oint_{C}\!\mathbf{-i\langle\mathbf{u}_{n\ \mathbf{k}}|\nabla_{\mathbf{k}}\mathbf{u}_{n\ \mathbf{k}}\rangle}\textrm{d}\mathbf{k} ,
\end{equation}
where $\mathbf{\textbf{u}}_{n\ \mathbf{k}}$ are the phonon eigen states. Combining with the first-principles-calculated phonon properties, the Berry phase of the nodal line along the $\Gamma$X is also $2\pi$. The summary of the Berry phases of acoustic nodal lines along other high-symmetry paths in Si, 4\textit{H}-SiC and \textit{c}-BN is listed in Table S1. For 4\textit{H}-SiC, it is found that the nodal rings and non-straight nodal lines consisting of two acoustic phonon branches exist in the BZ rather than straight nodal lines as displayed in Fig. S15. According to Ref. \cite{ubiquitous_silicon, zak_phase}, it is necessary to calculate the Zak's phase at the specific surface to determine whether topological phonons can be found. Here, we consider the (001) and (111) surfaces of Si, the (0001) surface of 4\textit{H}-SiC, and the (111) surface of \textit{c}-BN since these surfaces are relatively stable and the reconstructions are computationally feasible for the BTE calculation~\cite{bte}. Using the bulk harmonic IFCs from first-principles calculations and the Green's function with open-boundary conditions~\cite{surface_green_function}, the Zak's phase of different acoustic phonon branches across the whole BZ is calculated and plotted in Figs. S6, S8, S11 and S13, and the LDOS is plotted in Figs. S5, S7, S10 and S12. For the perfect surfaces of Si, 4\textit{H}-SiC, and c-BN, the topological surface phonons associated with acoustic phonon nodal lines can be allocated, and topological surface phonons with much lower frequencies can be observed. Such low-frequency topological surface phonons can contribute to thermal conductance and serve as multiple-phonon scattering channels for long wavelength phonons as well.

\begin{figure*}
    \centering
    \includegraphics[width=0.8\textwidth]{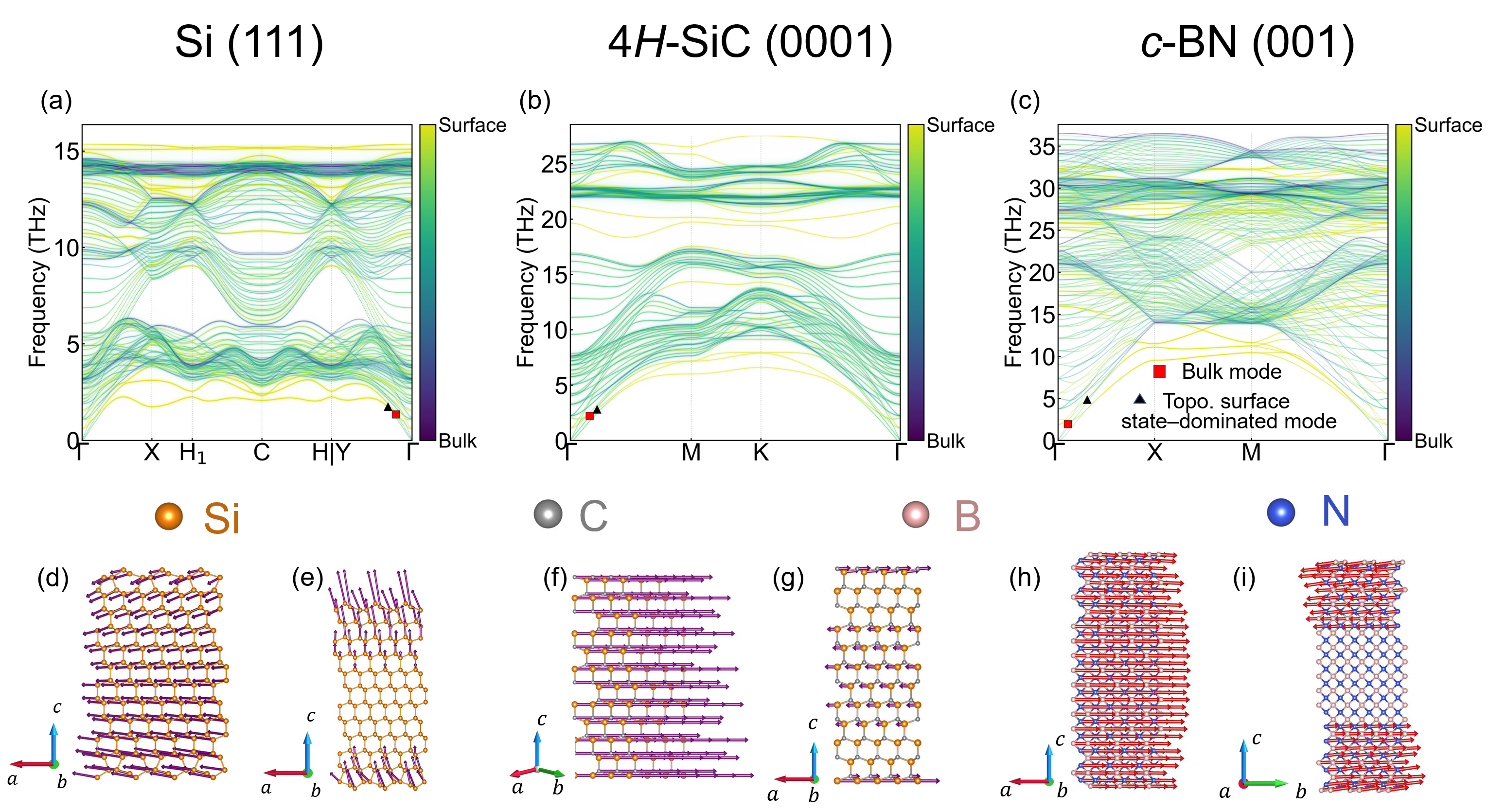}
    \caption{LDOS of the (a) Si (111), (b) 4\textit{H}-SiC (0001) and (c) \textit{c}-BN (001) thin films calculated from the NEPs, and the surface and bulk contributions to phonons are distinguished according to their amplitudes of vibrations of surface and bulk atoms. The selected atomic vibrations of typical bulk and topological surface state-dominated phonon modes for (d, e) Si (111), (f, g) 4\textit{H}-SiC (0001) and (h, i) \textit{c}-BN (001)  are plotted, respectively. Red squares represent for bulk modes, and black triangle for topological surface state-dominated modes.}
    \label{fig-3}
\end{figure*}

\subsection{Machine learning potentials for thin films}

In electronic applications, semiconductors always take the form of thin films, and surface structures undergo relaxations or reconstructions with much-reduced symmetry (see Fig. 1(f) for the ideal surface and 2$\times$1 reconstructed surface of Si (111) thin films). In addition, different surface types can emerge due to the nonequivalent atoms exposing to the vacuum. For example, Si thin films exhibit two types of surfaces (denoted as a type and b type, see Supporting Information Note S1.2 for more details) for the (001) and (111) orientations when cleaving from the bulk~\cite{ubiquitous_silicon}, although the primitive cell has two identical Si atoms. Phonon BTE with direct first-principles calculations is not applicable to calculate phonon transport properties of these thin films due to the increased computational cost. Here, we utilized machine learning potentials called NEPs with first-principles accuracy through the active learning strategy for thin films of Si, 4\textit{H}-SiC, and \textit{c}-BN (see Supporting Information for more details on training machine learning potentials) and calculated the phonon transport properties of thin films with different surface orientations. The energies and atomic forces of training sets of Si, 4\textit{H}-SiC, and \textit{c}-BN from DFT calculations and NEP predictions are plotted in Fig.~\ref{fig-2} (b-g), and good agreements are achieved for our NEPs. Other comparisons of phonon dispersion relations of semiconductor thin films between DFT and NEP calculations are shown in Fig. S20. Using these NEPs, we optimized the thin films with different surface orientations, and it was found that the (0001) surfaces of 4\textit{H}-SiC with no obvious reconstructions, while all the Si (001), Si (111) and \textit{c}-BN (111) surfaces show clear 2$\times$1 reconstructions, as displayed in Fig. S22, which are also consistent with previous experimental observations~\cite{si001_reconstruction, si111, bn_reconstruction}. Such surface reconstructions can lower the symmetry of thin films and lift the degeneracy of phonon modes including the low-frequency topological surface modes, compared with those of perfect surfaces (Figs.~S5, S7, and S12), which is also demonstrated by the phonon surface LDOS of Si and \textit{c}-BN thin films in Figs.~\ref{fig-3} (a), (c) and S19. On the contrary, the phonon LDOS of 4\textit{H}-SiC (0001) thin films shows consistency with those of perfect ones, as displayed by Figs.~\ref{fig-3} and S18. In the main text, we have displayed only Si (111) a-type thin film, 4\textit{H}-SiC (0001) thin film with terminations of C atoms, and \textit{c}-BN (001) a-type thin films with terminations of B atoms, denoted as Si (111), 4\textit{H}-SiC (0001), and \textit{c}-BN (001) thin films. Other types of semiconductor thin films can be found in the Supporting Information. Due to the bulk-boundary correspondence and the calculated Zak's phases at certain surfaces, low-frequency topological surface states associated with nontrivial acoustic phonon nodal lines can be found at all the considered surfaces. For Si (111) thin films, the phonon dispersion relations show large differences compared with those calculated from the Green's function (Figs. S5 and S8), especially for low-frequency topological surface states, which demonstrate wave-like curves along the high-symmetry paths due to the more severe surface reconstructions of (111) surfaces. For Si thin films with (001) a- and b-type surfaces, their surface reconstructions, the phonon dispersion relations and the low-frequency topological surface states demonstrate similar behaviors. For the \textit{c}-BN (001) thin films, the topological surface states undergo a slightly softening effect, implying possible large contributions to phonon transport.  For 4\textit{H}-SiC (0001) thin films, the low-frequency topological surface states are closely aligned with those obtained from the Green's function method. This agreement can be ascribed to the absence of significant surface reconstructions on these surfaces.

\begin{figure*}
    \centering
    \includegraphics[width=0.8\textwidth]{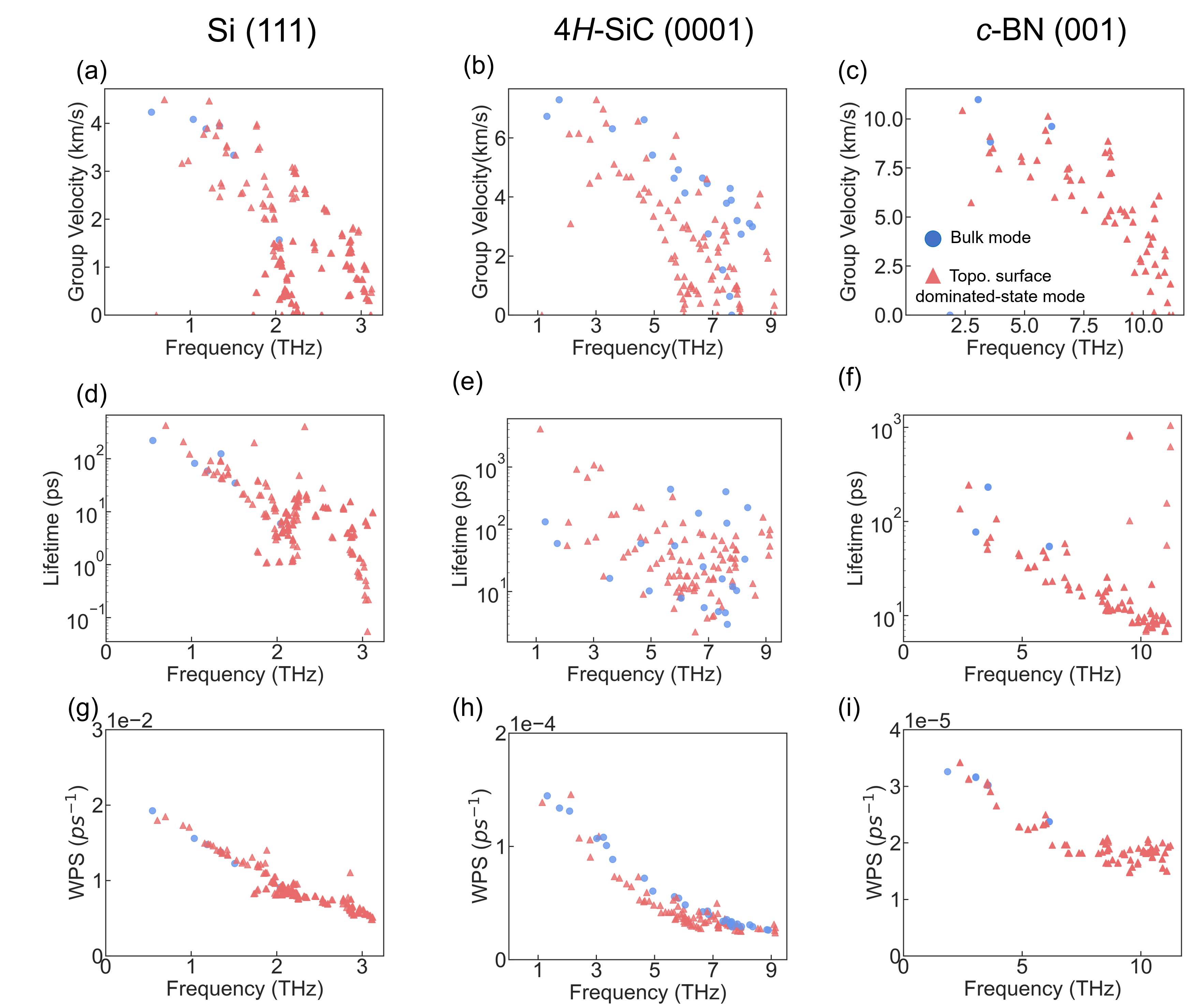}
    \caption{(a-c) Phonon group velocities, (d-f) lifetimes and (g-i) WPSs of the Si (111), 4\textit{H}-SiC (0001) and \textit{c}-BN (001) thin films with the thickness of  approximately 3 nm calculated from the BTE method and NEPs, respectively. The bulk and surface modes are marked separately.}
    \label{fig-4}
\end{figure*}

\subsection{Phonon transport properties and contributions of topological surface states}

Since topological surface states strongly couple with bulk states at their crossings, it becomes challenging to quantify the contribution of topological surface states. To distinguish between bulk and topological surface phonon states in the low-frequency regime, we calculated the phonon eigenvectors of the lowest four phonon branches and identified the topological surface state-dominated phonons based on atomic vibration amplitudes: a phonon is classified as topological surface state-dominated if the vibrational amplitude of surface atoms exceeds 1.2 times that of bulk atoms in its eigenvector. Typical position-dependent atomic vibrations of topological surface state-dominated and bulk phonons are demonstrated in Figs.~\ref{fig-3} (d-i) and S23. Accordingly, the phonon group velocity,  WPSs, lifetimes, and the resulting contribution to lattice thermal conductivity of low-frequency topological surface state-dominated phonons can be unambiguously extracted. In Fig.~\ref{fig-4}, we presented the phonon group velocities, WPSs, lifetimes for approximately 3-nm-thick Si (001), 4\textit{H}-SiC (0001), and \textit{c}-BN (111) thin films (results for other thin films can be found in Fig.~S21). Notably, many phonon modes are dominated by topological surface states, even those with long wavelengths. Phonons dominated by topological surface states of all semiconductor thin films demonstrate relatively high phonon group velocities and long lifetimes, suggesting a key role in heat conduction.

\begin{figure*}
    \centering
    \includegraphics[width=\textwidth]{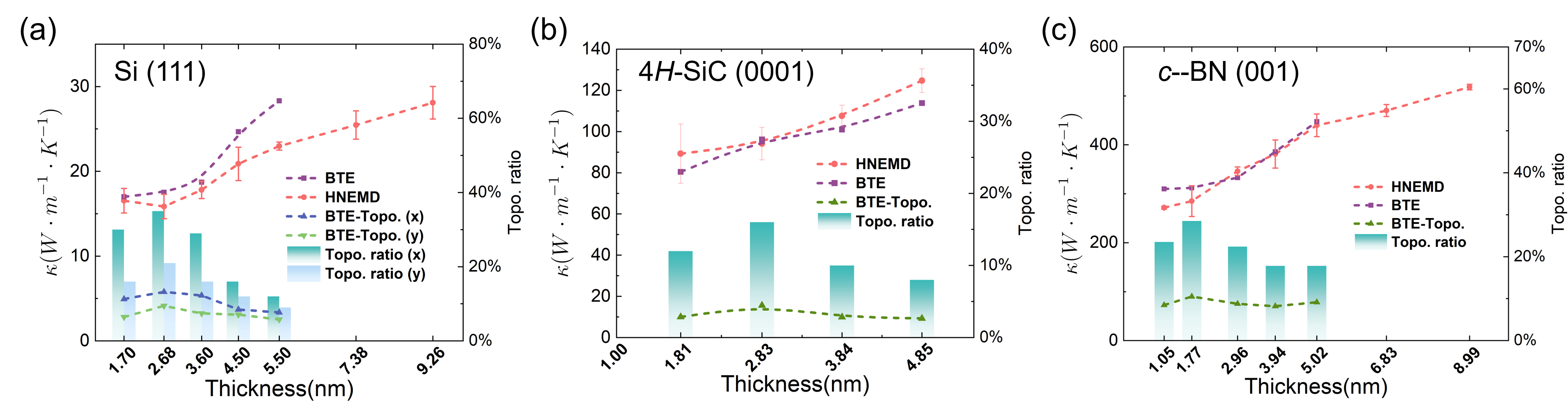}
    \caption{The thickness-dependent lattice thermal conductivities of (a)  Si (111) a-type, (b) 4\textit{H}-SiC (0001) and (c) \textit{c}-BN (001) thin films calculated from the BTE and HNEMD methods, and the absolute and relative contributions to lattice thermal conductivity of topological surface state-dominated phonons from the BTE are also extracted. Here, Topo.~represents topological surface state-dominated mode. }
    \label{fig-5}
\end{figure*}

The in-plane thermal conductivities of these thin films with the thickness from approximately 1 nm to 5 nm are further calculated using the NEPs with the BTE method and HNEMD simulations as shown in Fig.~\ref{fig-5}. The two methods yield good agreement where both are applicable. For thicker thin films, the large number of atoms in the unit cell makes BTE calculations computationally prohibitive; thus, only HNEMD is employed in those cases. Phonon transport in Si thin films exhibits pronounced surface-orientation dependence. For Si (001) a- and b-type thin films, the in-plane thermal conductivity follows a similar thickness-dependent trend: the thermal conductivity of a 1.4-nm-thickness thin film can be as high as 20$\%$ of the bulk value at 300 K ($\sim$150 W/m-K~\cite{Si_kappa}). In contrast, Si (111) thin films display lower thermal conductivity, with a 1.7-nm-thick film achieving only 12$\%$ of the bulk value at 300 K. For 4\textit{H}-SiC (0001) thin films, the in-plane thermal conductivity is relatively high (from 80 W/m-K to 115 W/m-K, or 17.0$\%$ to 24.4$\%$ of the in-plane bulk value of 471 W/m-K~\cite{SiC_kappa} as the thickness increases). Remarkably, \textit{c}-BN (001) thin films show the highest in-plane thermal conductivity from 300 W/m-K to 500 W/m-K (18.8$\%$ to 31$\%$ of the bulk value 1600 W/m-K~\cite{BN_kappa}) as the thickness increases, even at a thickness below 10 nm. Our results indicate that semiconductor thin films with atomically smooth surfaces can sustain relatively high thermal conductivity, even when their thickness is reduced below 10 nm.

More importantly, in the BTE method, the contribution of topological surface state-dominated phonons to thermal conductivity can be quantitatively extracted for all semiconductor thin films using the aforementioned identification criterion, as shown in Fig.~\ref{fig-5}. For Si thin films, as the thickness decreases, the contribution to the in-plane thermal conductivity of topological surface phonon states increases significantly. Specifically, for Si (111) thin films, topological surface state-dominated phonons contribute at least 30$\%$  of the total in-plane thermal conductivity when the thickness is below 4.5 nm in our calculations. Similarly, for Si (001) a- and b-type thin films, these phonons contribute no less than 20$\%$ when the thickness is below 3.02 nm. For c-BN (001) thin films, the contribution of topological surface state-dominated phonons accounts for more than 18$\%$ across all thicknesses in our calculations. Notably, their absolute contribution reaches as high as 82 W/m-K, reflecting the highest phonon group velocity of the topological surface state-dominated phonons among the three semiconductors. It is crucial to note that our calculations represent a lower bound for the contribution of low-frequency topological surface states to in-plane thermal transport. This is because we only accounted for phonon modes that are predominantly characterized by these topological states.

\begin{figure}
    \centering
    \includegraphics[width=\columnwidth]{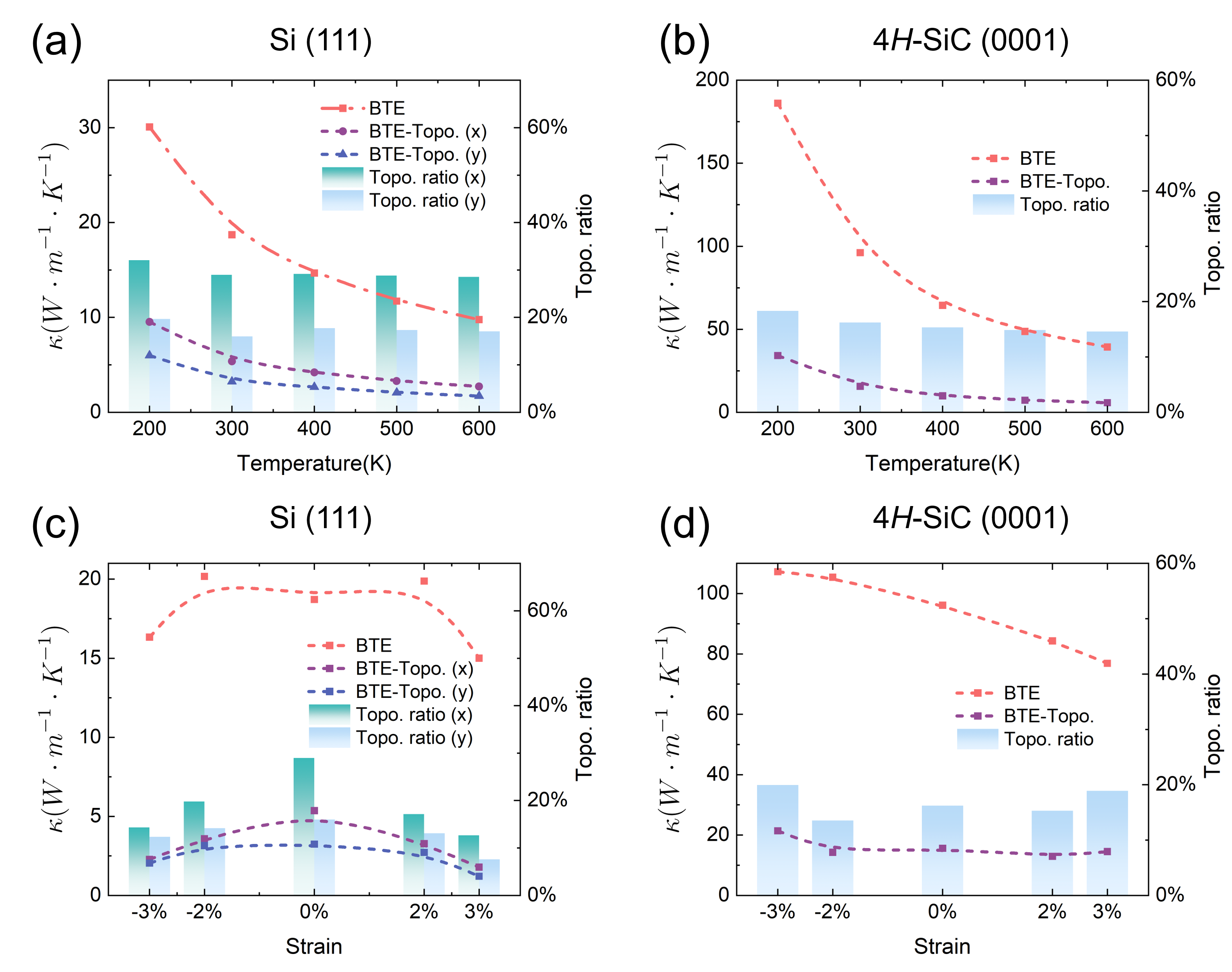}
    \caption{(a, b) Temperature-dependent and (c, d) strain-dependent room-temperature in-plane lattice thermal conductivity of Si (111) a-type and 4\textit{H}-SiC (0001) thin films with the thickness of approximately 3 nm. }
    \label{fig-6}
\end{figure}

We further investigated how temperature and biaxial strain modulate the in-plane phonon transport of low-frequency topological surface state-dominated phonons associated with the acoustic phonon nodal lines. We first calculated the in-plane thermal conductivity of Si (001) a-type and 4\textit{H}-SiC (0001) thin films using the BTE method and extracted the contributions of topological surface state-dominated phonons from 200 K to 600 K. The results are displayed in Figs.~\ref{fig-6} (a) and (b). The contribution of low-frequency topological surface state-dominated phonons to the total in-plane thermal conductivity follows the same temperature-dependent trend as the overall lattice thermal conductivity of both thin films. As temperature decreases, their absolute contribution slightly increases, which can be attributed to the enhancement of phonon lifetimes at lower temperatures (see Fig. S28). In Figs.~\ref{fig-6} (c) and (d), the strain-dependent in-plane thermal conductivity and the contributions of topological surface states of Si (001) a-type and 4\textit{H}-SiC (0001) thin films are displayed as the strain varies from -3$\%$ to 3$\%$. It was observed that for Si (111) thin films, the contribution of phonons dominated by topological surface states initially increases and then decreases under biaxial strain ranging from -3$\%$ to 3$\%$. In contrast, for 4\textit{H}-SiC (0001) thin films, the contribution of phonons dominated by topological surface states shows minimal variation across the same strain range. This discrepancy under biaxial strain can be attributed to the distinct surface characteristics: the Si (001) surfaces exhibit obvious structural reconstructions that are sensitive to strain and consequently modulate the low-frequency topological surface states, whereas the 4\textit{H}-SiC (0001) surfaces remain relatively pristine and structurally stable even under biaxial strain. 

\vspace{10pt}

\section{Conclusions}\label{sec4}

In summary, using machine learning potentials, we investigated the influence of low-frequency topological surface phonons, which arise due to the nontrivial acoustic phonon nodal lines, on the in-plane thermal transport in semiconductor thin films of Si, 4\textit{H}-SiC, and \textit{c}-BN with the thickness below 10 nm. By identifying low-frequency phonons dominated by topological surface states, their phonon group velocity, lifetimes, and contributions to the total in-plane thermal conductivity are unambiguously extracted. For Si (111) thin films, the low-frequency topological surface state-dominated phonons can account for at least 30$\%$ of the total in-plane thermal conductivity at 300 K, while for c-BN (001) thin films, their absolute contribution reaches as high as 82 W/m-K. Further calculations demonstrate that temperature and biaxial strains can modulate the transport behaviors of topological surface state-dominated phonons. With the evidence, we unveil the notable role of topological surface phonons in thermal transport in semiconductors and provide new physical insights into phonon transport at the nanoscale.

\section*{Acknowledgments}
This work was financially supported by the National Key R\&D Program of China (2022YFA1404400), the National Natural Science Foundation of China (Grant No. 12125504), the ``Hundred Talents Program'' of the Chinese Academy of Sciences, the Priority Academic Program Development (PAPD) of Jiangsu Higher Education Institutions, and the start-up funding of the Suzhou Institute for Advanced Research, University of Science and Technology of China. The numerical calculations in this paper have been done on the supercomputing system in the Supercomputing Center of University of Science and Technology of China. The authors also thank for the support of Open Source Supercomputing Center of S-A-I.

\section*{Conflicts of Interest}
The authors declare no conflicts of interest.


\bibliographystyle{elsarticle-num} 
\bibliography{arxiv}

@PREAMBLE{
 "\providecommand{\noopsort}[1]{}" 
 # "\providecommand{\singleletter}[1]{#1}%" 
}

@article{EleTopoCatalogue,
	title = {Catalogue of topological electronic materials},
	volume = {566},
	issn = {0028-0836, 1476-4687},
	doi = {10.1038/s41586-019-0944-6},
	number = {7745},
	journal = {Nature},
	author = {Zhang, Tiantian and Jiang, Yi and Song, Zhida and Huang, He and He, Yuqing and Fang, Zhong and Weng, Hongming and Fang, Chen},
	Month = {FEB 28},
	year = {2019},
	pages = {475--479}
}

@article{EleTopoCatalogue_2,
	title = {Comprehensive search for topological materials using symmetry indicators},
	volume = {566},
	issn = {0028-0836, 1476-4687},
	doi = {10.1038/s41586-019-0937-5},
	number = {7745},
	journal = {Nature},
	author = {Tang, Feng and Po, Hoi Chun and Vishwanath, Ashvin and Wan, Xiangang},
	month = feb,
	year = {2019},
	pages = {486--489},
}

@article{EleTopoCatalogue_3,
	title = {A complete catalogue of high-quality topological materials},
	volume = {566},
	issn = {1476-4687},
	doi = {10.1038/s41586-019-0954-4},
	number = {7745},
	journal = {Nature},
	author = {Vergniory, M. G. and Elcoro, L. and Felser, Claudia and Regnault, Nicolas and Bernevig, B. Andrei and Wang, Zhijun},
	month = feb,
	year = {2019},
	pages = {480--485}
}

@article{CatalogPhonon,
	title = {Catalog of topological phonon materials},
	volume = {384},
	issn = {0036-8075, 1095-9203},
	doi = {10.1126/science.adf8458},
	number = {6696},
	journal = {Science},
	author = {Xu, Yuanfeng and Vergniory, M. G. and Ma, Da-Shuai and Mañes, Juan L. and Song, Zhi-Da and Bernevig, B. Andrei and Regnault, Nicolas and Elcoro, Luis},
	month = may,
	year = {2024},
	pages = {eadf8458},
}

@article{KMModelTopo,
	title = {Z$_2$ topological order and the quantum spin hall effect},
	volume = {95},
	issn = {0031-9007, 1079-7114},
	doi = {10.1103/PhysRevLett.95.146802},
	number = {14},
	journal = {Physical Review Letters},
	author = {Kane, C. L. and Mele, E. J.},
	month = sep,
	year = {2005},
	pages = {146802},
}

@article{QuantumTopo_2007,
	title = {Quantum {spin} {hall} {insulator} {state} in {HgTe} {quantum} {wells}},
	volume = {318},
	issn = {0036-8075, 1095-9203},
	doi = {10.1126/science.1148047},	 
	number = {5851},
	journal = {Science},
	author = {K{\"o}nig, Markus and Wiedmann, Steffen and Br{\"u}ne, Christoph and Roth, Andreas and Buhmann, Hartmut and Molenkamp, Laurens W. and Qi, Xiao-Liang and Zhang, Shou-Cheng},
	month = nov,
	year = {2007},
	pages = {766--770},
}

@article{PhononBulkExp_2018,
	title = {Observation of {double} {Weyl} {phonons} in {parity}-{breaking} {FeSi}},
	volume = {121},
	issn = {0031-9007, 1079-7114},
	doi = {10.1103/PhysRevLett.121.035302},
	number = {3},
	journal = {Physical Review Letters},
	author = {Miao, H. and Zhang, T. T. and Wang, L. and Meyers, D. and Said, A. H. and Wang, Y. L. and Shi, Y. G. and Weng, H. M. and Fang, Z. and Dean, M. P. M.},
	month = jul,
	year = {2018},
	pages = {035302},
}

@article{GrapheneTopoExp_2023,
	title = {Direct {observation} of {topological} {phonons} in {graphene}},
	volume = {131},
	issn = {0031-9007, 1079-7114},
	doi = {10.1103/PhysRevLett.131.116602},
	number = {11},
	journal = {Physical Review Letters},
	author = {Li, Jiade and Li, Jiangxu and Tang, Jilin and Tao, Zhiyu and Xue, Siwei and Liu, Jiaxi and Peng, Hailin and Chen, Xing-Qiu and Guo, Jiandong and Zhu, Xuetao},
	month = sep,
	year = {2023},
	pages = {116602},
}

@article{PhononTopoSearc,
	title = {Computation and data driven discovery of topological phononic materials},
	volume = {12},
	issn = {2041-1723},
	doi = {10.1038/s41467-021-21293-2},	 
	number = {1},
	journal = {Nature Communications},
	author = {Li, Jiangxu and Liu, Jiaxi and Baronett, Stanley A. and Liu, Mingfeng and Wang, Lei and Li, Ronghan and Chen, Yun and Li, Dianzhong and Zhu, Qiang and Chen, Xing-Qiu},
	month = feb,
	year = {2021},
	pages = {1204},
}

@article{TopoQuanChem_2017,
	title = {Topological quantum chemistry},
	volume = {547},
	issn = {1476-4687},
	doi = {10.1038/nature23268},	 
	number = {7663},
	journal = {Nature},
	author = {Bradlyn, Barry and Elcoro, L. and Cano, Jennifer and Vergniory, M. G. and Wang, Zhijun and Felser, C. and Aroyo, M. I. and Bernevig, B. Andrei},
	month = jul,
	year = {2017},
	pages = {298--305},
}

@article{SymIndicator,
	title = {Symmetry-based indicators of band topology in the 230 space groups},
	volume = {8},
	issn = {2041-1723},
	doi = {10.1038/s41467-017-00133-2},
	number = {1},
	journal = {Nature Communications},
	author = {Po, Hoi Chun and Vishwanath, Ashvin and Watanabe, Haruki},
	month = jun,
	year = {2017},
	pages = {50},
}

@article{surface_conduction_2012,
	title = {Surface conduction of topological {Dirac} electrons in bulk insulating {$Bi_2Se_3$}},
	volume = {8},
	issn = {1745-2473, 1745-2481},
	doi = {10.1038/nphys2286},
	language = {en},
	number = {6},
	journal = {Nature Physics},
	author = {Kim, Dohun and Cho, Sungjae and Butch, Nicholas P. and Syers, Paul and Kirshenbaum, Kevin and Adam, Shaffique and Paglione, Johnpierre and Fuhrer, Michael S.},
	month = jun,
	year = {2012},
	pages = {459--463},
}

@article{tuning_surface_conduction_2015,
	title = {Tuning {bulk} and {surface} {conduction} in the {proposed} {topological} {Kondo} {insulator} {$SmB_6$}},
	volume = {114},
	issn = {0031-9007, 1079-7114},
	doi = {10.1103/PhysRevLett.114.096601},
	language = {en},
	number = {9},
	journal = {Physical Review Letters},
	author = {Syers, Paul and Kim, Dohun and Fuhrer, Michael S. and Paglione, Johnpierre},
	month = mar,
	year = {2015},
	pages = {096601},
}

@article{nanoscale_ele_transport_2016,
	title = {Nanoscale electron transport at the surface of a topological insulator},
	volume = {7},
	issn = {2041-1723},
	doi = {10.1038/ncomms11381},
	language = {en},
	number = {1},
	journal = {Nature Communications},
	author = {Bauer, Sebastian and Bobisch, Christian A.},
	month = apr,
	year = {2016},
	pages = {11381},
}

@article{suface_diode_effect_2025,
	title = {Realizing a topological diode effect on the surface of a topological {Kondo} insulator},
	volume = {122},
	issn = {0027-8424, 1091-6490},
	doi = {10.1073/pnas.2417709122},
	language = {en},
	number = {12},
	journal = {Proceedings of the National Academy of Sciences},
	author = {Zhang, Jiawen and Hua, Zhenqi and Wang, Chengwei and Smidman, Michael and Graf, David and Thomas, Sean and Rosa, Priscila F. S. and Wirth, Steffen and Dai, Xi and Xiong, Peng and Yuan, Huiqiu and Wang, Xiaoyu and Jiao, Lin},
	month = mar,
	year = {2025},
	pages = {e2417709122},
}

@article{topo_phonon_thermoelectric_2024,
	title = {Topological {phonons} and {thermoelectric} {conversion} in {crystalline} {materials}},
	volume = {34},
	issn = {1616-301X, 1616-3028},
	doi = {10.1002/adfm.202401684},
	language = {en},
	number = {33},
	journal = {Advanced Functional Materials},
	author = {Ding, Zhong‐Ke and Zeng, Yu‐Jia and Liu, Wangping and Tang, Li‐Ming and Chen, Ke‐Qiu},
	month = aug,
	year = {2024},
	pages = {2401684},
}

@article{surface_reconst_Si,
	title = {Role of {electronic} {correlation} in the {Si}(100) {reconstruction}: {A} {quantum} {monte} {carlo} {study}},
	volume = {87},
	issn = {0031-9007, 1079-7114},
	shorttitle = {Role of {Electronic} {Correlation} in the {Si}(100) {Reconstruction}},
	doi = {10.1103/PhysRevLett.87.016105},
	language = {en},
	number = {1},
	journal = {Physical Review Letters},
	author = {Healy, Sorcha B. and Filippi, Claudia and Kratzer, P. and Penev, E. and Scheffler, M.},
	month = jun,
	year = {2001},
	pages = {016105},
}

@article{surface_reconst_GaN,
	title = {{GaN}(0001) surface states: {Experimental} and theoretical fingerprints to identify surface reconstructions},
	volume = {88},
	issn = {1098-0121, 1550-235X},
	shorttitle = {{GaN}(0001) surface states},
	doi = {10.1103/PhysRevB.88.125304},
	language = {en},
	number = {12},
	journal = {Physical Review B},
	author = {Himmerlich, M. and Lymperakis, L. and Gutt, R. and Lorenz, P. and Neugebauer, J. and Krischok, S.},
	month = sep,
	year = {2013},
	pages = {125304},
}

@article{dftb_Si_thin_films,
	title = {Direct first-principle-based study of mode-wise in-plane phonon transport in ultrathin silicon films},
	volume = {143},
	issn = {00179310},
	doi = {10.1016/j.ijheatmasstransfer.2019.118507},
	language = {en},
	journal = {International Journal of Heat and Mass Transfer},
	author = {Wang, Qi and Guo, Ruiqiang and Chi, Cheng and Zhang, Kai and Huang, Baoling},
	month = nov,
	year = {2019},
	pages = {118507},
}

@article{Si_thin_film_exp,
	title = {Tuning {thermal} {transport} in {ultrathin} {silicon} {membranes} by {surface} {nanoscale} {engineering}},
	volume = {9},
	issn = {1936-0851, 1936-086X},
	doi = {10.1021/nn506792d},
	language = {en},
	number = {4},
	journal = {ACS Nano},
	author = {Neogi, Sanghamitra and Reparaz, J. Sebastian and Pereira, Luiz Felipe C. and Graczykowski, Bartlomiej and Wagner, Markus R. and Sledzinska, Marianna and Shchepetov, Andrey and Prunnila, Mika and Ahopelto, Jouni and Sotomayor-Torres, Clivia M. and Donadio, Davide},
	month = apr,
	year = {2015},
	pages = {3820--3828},
}

@article{review_surface_cal_2018,
	title = {A review of computational phononics: the bulk, interfaces, and surfaces},
	volume = {53},
	issn = {0022-2461, 1573-4803},
	shorttitle = {A review of computational phononics},
	doi = {10.1007/s10853-017-1728-8},
	language = {en},
	number = {8},
	journal = {Journal of Materials Science},
	author = {VanGessel, Francis and Peng, Jie and Chung, Peter W.},
	month = apr,
	year = {2018},
	pages = {5641--5683},
}

@article{S_matrix_graphene,
	title = {Robustness and scattering behavior of topological phonons in crystalline materials},
	volume = {109},
	issn = {2469-9950, 2469-9969},
	doi = {10.1103/PhysRevB.109.245104},
	language = {en},
	number = {24},
	journal = {Physical Review B},
	author = {Ding, Zhong-Ke and Zeng, Yu-Jia and Pan, Hui and Luo, Nannan and Tang, Li-Ming and Zeng, Jiang and Chen, Ke-Qiu},
	month = jun,
	year = {2024},
	pages = {245104},
}

@article{s_matrix_corner_state,
	title = {Phonon transport of higher-order topological states in {$MoTe_2$}  and {$WTe_2$}},
	volume = {112},
	issn = {2469-9950, 2469-9969},
	doi = {10.1103/ctth-l8bx},
	language = {en},
	number = {7},
	journal = {Physical Review B},
	author = {Liu, Wangping and Ding, Zhong-Ke and He, Ran and Ding, Changhao and Yao, Yuan and Luo, Nannan and Zeng, Jiang and Tang, Li-Ming and Chen, Ke-Qiu},
	month = aug,
	year = {2025},
	pages = {075415},
}

@article{topo_phonon_transport_bulk,
    author = {Luo, Xiaobing and Zhang, Haopeng and Chen, Peng and Yan, Yanci and Wu, Hong and Wang, Xiaotian and Ding, Guangqian},
    title = {Insights into the lattice thermal conductivity of solids carrying type-I and type-II Weyl point phonons},
    journal = {Journal of Applied Physics},
    volume = {138},
    number = {12},
    pages = {125102},
    year = {2025},
    month = {09},
    issn = {0021-8979},
    doi = {10.1063/5.0287221},
}

@article{topo_ele_transport_surface,
author = {Brahlek, Matthew},
title = {Criteria for realizing room-temperature electrical transport applications of topological materials},
journal = {Advanced Materials},
volume = {32},
number = {50},
pages = {2005698},
doi = {10.1002/adma.202005698},
year = {2020}
}

@article{phonon_NL_PRB,
	title = {Symmetry-enforced straight nodal-line phonons},
	volume = {104},
	issn = {2469-9950, 2469-9969},
	doi = {10.1103/PhysRevB.104.024304},
	language = {en},
	number = {2},
	journal = {Physical Review B},
	author = {Liu, Guang and Jin, Yuanjun and Chen, Zhongjia and Xu, Hu},
	month = jul,
	year = {2021},
	pages = {024304},
}

@article{phonon_NL_review,
    author = {Wang, Xiaotian and Yang, Tie and Cheng, Zhenxiang and Surucu, Gokhan and Wang, Jianhua and Zhou, Feng and Zhang, Zeying and Zhang, Gang},
    title = {Topological nodal line phonons: Recent advances in materials realization},
    journal = {Applied Physics Reviews},
    volume = {9},
    number = {4},
    pages = {041304},
    year = {2022},
    month = {11},
    issn = {1931-9401},
    doi = {10.1063/5.0095281},
}

@article{phonon_NL_NS,
	title = {Hybrid nodal surface and nodal line phonons in solids},
	volume = {108},
	issn = {2469-9950, 2469-9969},
	doi = {10.1103/PhysRevB.108.115153},
	language = {en},
	number = {11},
	journal = {Physical Review B},
	author = {Dong, Wen-Han and Pan, Jinbo and Sun, Jia-Tao and Du, Shixuan},
	month = sep,
	year = {2023},
	pages = {115153},
}

@article{ubiquitous_silicon,
	title = {Ubiquitous {topological} {states} of {phonons} in {solids}: {Silicon} as a {model} {material}},
	volume = {22},
	issn = {1530-6984, 1530-6992},
	shorttitle = {Ubiquitous {Topological} {States} of {Phonons} in {Solids}},
	doi = {10.1021/acs.nanolett.1c04299},
	language = {en},
	number = {5},
	journal = {Nano Letters},
	author = {Liu, Yizhou and Zou, Nianlong and Zhao, Sibo and Chen, Xiaobin and Xu, Yong and Duan, Wenhui},
	month = mar,
	year = {2022},
	pages = {2120--2126},
}

@article{phonon_NL_3d,
	title = {Topological {phonons} and {Weyl} {lines} in {three} {dimensions}},
	volume = {117},
	issn = {0031-9007, 1079-7114},
	doi = {10.1103/PhysRevLett.117.068001},
	language = {en},
	number = {6},
	journal = {Physical Review Letters},
	author = {Stenull, Olaf and Kane, C. L. and Lubensky, T. C.},
	month = aug,
	year = {2016},
	pages = {068001},
}

@article{data_search_phonon_topo,
	title = {Computation and data driven discovery of topological phononic materials},
	volume = {12},
	issn = {2041-1723},
	doi = {10.1038/s41467-021-21293-2},
	language = {en},
	number = {1},
	journal = {Nature Communications},
	author = {Li, Jiangxu and Liu, Jiaxi and Baronett, Stanley A. and Liu, Mingfeng and Wang, Lei and Li, Ronghan and Chen, Yun and Li, Dianzhong and Zhu, Qiang and Chen, Xing-Qiu},
	month = feb,
	year = {2021},
	pages = {1204},
}

@article{phonon_NL_MoB2,
	title = {Phononic {helical} {nodal} {lines} with {PT} {protection} in {$MoB_2$} },
	volume = {123},
	issn = {0031-9007, 1079-7114},
	doi = {10.1103/PhysRevLett.123.245302},
	language = {en},
	number = {24},
	journal = {Physical Review Letters},
	author = {Zhang, T. T. and Miao, H. and Wang, Q. and Lin, J. Q. and Cao, Y. and Fabbris, G. and Said, A. H. and Liu, X. and Lei, H. C. and Fang, Z. and Weng, H. M. and Dean, M. P. M.},
	month = dec,
	year = {2019},
	pages = {245302},
}

@article{Si_dftb,
	title = {Direct first-principle-based study of mode-wise in-plane phonon transport in ultrathin silicon films},
	volume = {143},
	issn = {00179310},
	doi = {10.1016/j.ijheatmasstransfer.2019.118507},
	language = {en},
	journal = {International Journal of Heat and Mass Transfer},
	author = {Wang, Qi and Guo, Ruiqiang and Chi, Cheng and Zhang, Kai and Huang, Baoling},
	month = nov,
	year = {2019},
	pages = {118507},
}

@article{SiC_dft,
	title = {Phonon thermal transport in {2H}, {4H} and {6H} silicon carbide from first principles},
	volume = {1},
	issn = {25425293},
	doi = {10.1016/j.mtphys.2017.05.004},
	language = {en},
	journal = {Materials Today Physics},
	author = {Protik, Nakib Haider and Katre, Ankita and Lindsay, Lucas and Carrete, Jesús and Mingo, Natalio and Broido, David},
	month = jun,
	year = {2017},
	pages = {31--38},
}

@article{GaN_dft,
	title = {First-principles study on thermal transport properties of {GaN} under different cross-plane strain},
	volume = {233},
	issn = {00179310},
	doi = {10.1016/j.ijheatmasstransfer.2024.126049},
	language = {en},
	journal = {International Journal of Heat and Mass Transfer},
	author = {Xue, Juan and Li, Fengyi and Fan, Aoran and Ma, Weigang and Zhang, Xing},
	month = nov,
	year = {2024},
	pages = {126049},
}

@article{Si_exp_dft,
	title = {Temperature-dependent phonon lifetimes and thermal conductivity of silicon by inelastic neutron scattering and \textit{ab initio} calculations},
	volume = {102},
	issn = {2469-9950, 2469-9969},
	doi = {10.1103/PhysRevB.102.174311},
	language = {en},
	number = {17},
	journal = {Physical Review B},
	author = {Kim, D. S. and Hellman, O. and Shulumba, N. and Saunders, C. N. and Lin, J. Y. Y. and Smith, H. L. and Herriman, J. E. and Niedziela, J. L. and Abernathy, D. L. and Li, C. W. and Fultz, B.},
	month = nov,
	year = {2020},
	pages = {174311},
}

@article{nanoscale_thermal,
	title = {Nanoscale thermal transport},
	volume = {93},
	issn = {0021-8979, 1089-7550},
	doi = {10.1063/1.1524305},
	language = {en},
	number = {2},
	journal = {Journal of Applied Physics},
	author = {Cahill, David G. and Ford, Wayne K. and Goodson, Kenneth E. and Mahan, Gerald D. and Majumdar, Arun and Maris, Humphrey J. and Merlin, Roberto and Phillpot, Simon R.},
	month = jan,
	year = {2003},
	pages = {793--818},
}

@article{Si_kappa,
	title = {Thermal {conductivity} of {silicon} and {germanium} from 3°{K} to the {melting} {point}},
	volume = {134},
	issn = {0031-899X},
	doi = {10.1103/PhysRev.134.A1058},
	language = {en},
	number = {4A},
	journal = {Physical Review},
	author = {Glassbrenner, C. J. and Slack, Glen A.},
	month = may,
	year = {1964},
	pages = {A1058--A1069},
}

@article{SiC_kappa,
	title = {Anisotropic thermal conductivity of {4H} and {6H} silicon carbide measured using time-domain thermoreflectance},
	volume = {3},
	issn = {25425293},
	doi = {10.1016/j.mtphys.2017.12.005},
	language = {en},
	journal = {Materials Today Physics},
	author = {Qian, Xin and Jiang, Puqing and Yang, Ronggui},
	month = dec,
	year = {2017},
	pages = {70--75},
}

@article{BN_kappa,
	title = {Ultrahigh thermal conductivity in isotope-enriched cubic boron nitride},
	volume = {367},
	issn = {0036-8075, 1095-9203},
	doi = {10.1126/science.aaz6149},
	language = {en},
	number = {6477},
	journal = {Science},
	author = {Chen, Ke and Song, Bai and Ravichandran, Navaneetha K. and Zheng, Qiye and Chen, Xi and Lee, Hwijong and Sun, Haoran and Li, Sheng and Udalamatta Gamage, Geethal Amila Gamage and Tian, Fei and Ding, Zhiwei and Song, Qichen and Rai, Akash and Wu, Hanlin and Koirala, Pawan and Schmidt, Aaron J. and Watanabe, Kenji and Lv, Bing and Ren, Zhifeng and Shi, Li and Cahill, David G. and Taniguchi, Takashi and Broido, David and Chen, Gang},
	month = jan,
	year = {2020},
	pages = {555--559},
}

@article{zak_phase,
	title = {Topological origin of antichiral edge states induced by a nonchiral phonon},
	volume = {109},
	issn = {2469-9950, 2469-9969},
	doi = {10.1103/PhysRevB.109.155410},
	language = {en},
	number = {15},
	journal = {Physical Review B},
	author = {Su, Yunlong and Li, Gang},
	month = apr,
	year = {2024},
	pages = {155410},
}

@article{surface_green_function,
	title = {Highly convergent schemes for the calculation of bulk and surface {Green} functions},
	volume = {15},
	issn = {0305-4608},
	doi = {10.1088/0305-4608/15/4/009},
	number = {4},
	journal = {Journal of Physics F: Metal Physics},
	author = {Sancho, M P Lopez and Sancho, J M Lopez and Sancho, J M L and Rubio, J},
	month = apr,
	year = {1985},
	pages = {851--858},
}

@article{bte,
	title = {Intrinsic lattice thermal conductivity of semiconductors from first principles},
	volume = {91},
	issn = {0003-6951, 1077-3118},
	doi = {10.1063/1.2822891},
	language = {en},
	number = {23},
	journal = {Applied Physics Letters},
	author = {Broido, D. A. and Malorny, M. and Birner, G. and Mingo, Natalio and Stewart, D. A.},
	month = dec,
	year = {2007},
	pages = {231922},
}

@article{si111,
	title = {{$\pi$}-bonded molecular and chain models for the {Si}(111) surface},
	volume = {26},
	issn = {0163-1829},
	doi = {10.1103/PhysRevB.26.4762},
	language = {en},
	number = {8},
	journal = {Physical Review B},
	author = {Chadi, D. J.},
	month = oct,
	year = {1982},
	pages = {4762--4765},
}

@article{bn_reconstruction,
	title = {First-principles study on energetics of {cBN}(001) reconstructed surfaces},
	volume = {341},
	issn = {00396028},
	doi = {10.1016/0039-6028(95)00826-8},
	language = {en},
	number = {3},
	journal = {Surface Science},
	author = {Yamauchi, Jun and Tsukada, Masaru and Watanabe, Satoshi and Sugino, Osamu},
	month = nov,
	year = {1995},
	pages = {L1037--L1041},
}

@article{si001_reconstruction,
	title = {Room temperature {Si}(001)-(2 × 1) reconstruction solved by {X}-ray diffraction},
	volume = {375},
	issn = {0039-6028},
	doi = {10.1016/S0039-6028(97)80005-2},
	number = {1},
	journal = {Surface Science},
	author = {Felici, R. and Robinson, I. K. and Ottaviani, C. and Imperatori, P. and Eng, P. and Perfetti, P.},
	month = mar,
	year = {1997},
	pages = {55--62},
}

@article{phonopy,
	title = {First principles phonon calculations in materials science},
	volume = {108},
	issn = {13596462},
	doi = {10.1016/j.scriptamat.2015.07.021},
	language = {en},
	journal = {Scripta Materialia},
	author = {Togo, Atsushi and Tanaka, Isao},
	month = nov,
	year = {2015},
	pages = {1--5},
}

@article{PAW_dft,
	title = {Generalized {gradient} {approximation} {made} {simple}},
	volume = {77},
	issn = {0031-9007, 1079-7114},
	doi = {10.1103/PhysRevLett.77.3865},
	language = {en},
	number = {18},
	journal = {Physical Review Letters},
	author = {Perdew, John P. and Burke, Kieron and Ernzerhof, Matthias},
	month = oct,
	year = {1996},
	pages = {3865--3868},
}

@article{PBEsol,
	title = {Restoring the {density}-{gradient} {expansion} for {exchange} in {solids} and {surfaces}},
	volume = {100},
	issn = {0031-9007, 1079-7114},
	doi = {10.1103/PhysRevLett.100.136406},
	language = {en},
	number = {13},
	journal = {Physical Review Letters},
	author = {Perdew, John P. and Ruzsinszky, Adrienn and Csonka, Gábor I. and Vydrov, Oleg A. and Scuseria, Gustavo E. and Constantin, Lucian A. and Zhou, Xiaolan and Burke, Kieron},
	month = apr,
	year = {2008},
	pages = {136406},
}

@article{revPBE,
	title = {Comment on “{generalized} {gradient} {approximation} {made} {simple}”},
	volume = {80},
	issn = {0031-9007, 1079-7114},
	doi = {10.1103/PhysRevLett.80.890},
	language = {en},
	number = {4},
	journal = {Physical Review Letters},
	author = {Zhang, Yingkai and Yang, Weitao},
	month = jan,
	year = {1998},
	pages = {890--890},
}

@article{vasp,
	title = {Efficiency of ab-initio total energy calculations for metals and semiconductors using a plane-wave basis set},
	volume = {6},
	issn = {09270256},
	doi = {10.1016/0927-0256(96)00008-0},
	language = {en},
	number = {1},
	journal = {Computational Materials Science},
	author = {Kresse, G. and Furthmüller, J.},
	month = jul,
	year = {1996},
	pages = {15--50},
}

@article{kseq,
	title = {Self-{consistent} {equations} {including} {exchange} and {correlation} {effects}},
	volume = {140},
	issn = {0031-899X},
	doi = {10.1103/PhysRev.140.A1133},
	language = {en},
	number = {4A},
	journal = {Physical Review},
	author = {Kohn, W. and Sham, L. J.},
	month = nov,
	year = {1965},
	pages = {A1133--A1138},
}

@article{plane_wave,
	title = {Efficient iterative schemes for \textit{ab initio} total-energy calculations using a plane-wave basis set},
	volume = {54},
	issn = {0163-1829, 1095-3795},
	doi = {10.1103/PhysRevB.54.11169},
	language = {en},
	number = {16},
	journal = {Physical Review B},
	author = {Kresse, G. and Furthmüller, J.},
	month = oct,
	year = {1996},
	pages = {11169--11186},
}

@article{nepactive,
	title = {{MLIP}-3: {Active} learning on atomic environments with moment tensor potentials},
	volume = {159},
	issn = {0021-9606, 1089-7690},
	shorttitle = {{MLIP}-3},
	doi = {10.1063/5.0155887},
	language = {en},
	number = {8},
	journal = {The Journal of Chemical Physics},
	author = {Podryabinkin, Evgeny and Garifullin, Kamil and Shapeev, Alexander and Novikov, Ivan},
	month = aug,
	year = {2023},
	pages = {084112},
}

@article{hnemd,
	title = {Homogeneous nonequilibrium molecular dynamics method for heat transport and spectral decomposition with many-body potentials},
	volume = {99},
	issn = {2469-9950, 2469-9969},
	doi = {10.1103/PhysRevB.99.064308},
	language = {en},
	number = {6},
	journal = {Physical Review B},
	author = {Fan, Zheyong and Dong, Haikuan and Harju, Ari and Ala-Nissila, Tapio},
	month = feb,
	year = {2019},
	pages = {064308},
}

@article{shengbte,
	title = {{ShengBTE}: {A} solver of the {Boltzmann} transport equation for phonons},
	volume = {185},
	issn = {00104655},
	shorttitle = {{ShengBTE}},
	doi = {10.1016/j.cpc.2014.02.015},
	language = {en},
	number = {6},
	journal = {Computer Physics Communications},
	author = {Li, Wu and Carrete, Jesús and A. Katcho, Nebil and Mingo, Natalio},
	month = jun,
	year = {2014},
	pages = {1747--1758},
}

@article{gpumd_4.0,
	title = {{GPUMD} 4.0: {A} high‐performance molecular dynamics package for versatile materials simulations with machine‐learned potentials},
	volume = {3},
	issn = {2940-9489, 2940-9497},
	shorttitle = {{GPUMD} 4.0},
	doi = {10.1002/mgea.70028},
	language = {en},
	number = {3},
	journal = {Materials Genome Engineering Advances},
	author = {Xu, Ke and Bu, Hekai and Pan, Shuning and Lindgren, Eric and Wu, Yongchao and Wang, Yong and Liu, Jiahui and Song, Keke and Xu, Bin and Li, Yifan and Hainer, Tobias and Svensson, Lucas and Wiktor, Julia and Zhao, Rui and Huang, Hongfu and Qian, Cheng and Zhang, Shuo and Zeng, Zezhu and Zhang, Bohan and Tang, Benrui and Xiao, Yang and Yan, Zihan and Shi, Jiuyang and Liang, Zhixin and Wang, Junjie and Liang, Ting and Cao, Shuo and Wang, Yanzhou and Ying, Penghua and Xu, Nan and Chen, Chengbing and Zhang, Yuwen and Chen, Zherui and Wu, Xin and Jiang, Wenwu and Berger, Esme and Li, Yanlong and Chen, Shunda and Gabourie, Alexander J. and Dong, Haikuan and Xiong, Shiyun and Wei, Ning and Chen, Yue and Xu, Jianbin and Ding, Feng and Sun, Zhimei and Ala‐Nissila, Tapio and Harju, Ari and Zheng, Jincheng and Guan, Pengfei and Erhart, Paul and Sun, Jian and Ouyang, Wengen and Su, Yanjing and Fan, Zheyong},
	month = sep,
	year = {2025},
	pages = {e70028},
}

@article{SiC_SOC,
	title = {First-principles study on reconstruction of {4H}-{SiC}(0001) and ({$000\overline{1}$})},
	volume = {647},
	issn = {00396028},
	doi = {10.1016/j.susc.2015.11.019},
	language = {en},
	journal = {Surface Science},
	author = {Kaneko, Tomoaki and Yamasaki, Takahiro and Tajima, Nobuo and Ohno, Takahisa},
	month = may,
	year = {2016},
	pages = {45--50},
}

@article{wieder_topological_2021,
	title = {Topological materials discovery from crystal symmetry},
	volume = {7},
	issn = {2058-8437},
	doi = {10.1038/s41578-021-00380-2},
	language = {en},
	number = {3},
	journal = {Nature Reviews Materials},
	author = {Wieder, Benjamin J. and Bradlyn, Barry and Cano, Jennifer and Wang, Zhijun and Vergniory, Maia G. and Elcoro, Luis and Soluyanov, Alexey A. and Felser, Claudia and Neupert, Titus and Regnault, Nicolas and Bernevig, B. Andrei},
	month = nov,
	year = {2021},
	pages = {196--216},
}

@misc{zhang_new_2025,
	title = {New {Advances} in {Phonons}: {From} {Band} {Topology} to {Quasiparticle} {Chirality}},
	shorttitle = {New {Advances} in {Phonons}},
	doi = {10.48550/arXiv.2505.06179},
	publisher = {arXiv},
	author = {Zhang, Tiantian and Liu, Yizhou and Miao, Hu and Murakami, Shuichi},
	month = may,
	year = {2025},
	note = {arXiv:2505.06179},
}

\end{document}